\documentstyle[12pt]{article}
\begin{document}
\newcommand{\be}{\begin{equation}}
\newcommand{\ee}{\end{equation}}
\newcommand{\ba}{\begin{eqnarray}}
\newcommand{\ea}{\end{eqnarray}}
\begin{flushright}
{\Large IHEP-95-139\\
December 1995}
\end{flushright}
\vspace{3cm}
\begin{center}
{\Large\bf Asymptotic solution of Schwinger -- Dyson equation
for the gluon propagator in the infrared region
\footnotemark}\\
\footnotetext{Talk presented at the Infernational Conference
on High Energy Physics (Diffraction -- 95), September 6 -- 11,
1995, Novyi Svet, Crimea}

\vspace{0.2cm}

{\bf A.I. Alekseev}\\

\vspace{0.2cm}

{\small {\it Institute for High Energy Physics, Protvino 142284,
Moscow Region, Russia}\\
E-mail address: alekseev@mx.ihep.su\\}

\vspace{3cm}

\end{center}

The equation for the gluon propagator in the approach of
Baker-Ball-Zachariasen is  considered. The possibility of
non-integer power infrared behaviour is studied,
$D(q) \sim (q^2)^{-c}$, $q^2 \rightarrow 0$.
It is shown that  the characteristic equation for the exponent
has no solutions at
$-1\leq c\leq 3$.
The approximations made to obtain the closed integral equation
are analysed and the conclusion on the infrared behaviour of
the gluon propagator
 $D(q) \sim 1/(q^2)^2$, $q^2 \rightarrow 0$
is made when the transverse part of the triple gluon vertex
is taken into account.

\newpage

The study of the infrared behaviour of  Green's functions
without any doubt is an important problem in QCD.
We mean first of all the infrared behaviour of the gluon
propagator. This behaviour is considered to have
a decisive role in the problem of confinement,
calculation of vacuum expectation values of gluon and quark fields,
the description of spontaneous chiral symmetry
breaking, etc.~\cite{1,2}.
The renormalization group method allows one to obtain the
asymptotic behaviour of the gluon propagator at large
momentum~\cite{3}.
As  we consider the infrared behaviour, then the approximate
solution of the exact Schwinger -- Dyson equations for Green's
functions seems to be the most adequate method of nonperturbative
nonlattice study.
At present the commonly accepted result on the character
of the infrared behaviour of the gluon propagator is absent
in the literature.

Nonuniqueness of the solutions can be conditioned by
nonlinearity of the corresponding integral equations
as well as the difference of the truncation methods
of the equations for Green's functions.
As it is shown in Refs.~\cite{4},
when finding nonperturbative solutions
(and the infrared region is completely nonperturbative one)
the problem of supplementary boundary conditions is of
great importance.
In Refs.~[5-9] the gluon propagator behaviour
 $D(q)\sim 1/(q^2)^2, q^2\rightarrow 0$ was obtained.
This behaviour gives a linear confining $q \bar q$
potential at large distances in the Born approximation.
Moreover this behaviour serves as a basis for physically
attractive picture where  the QCD vacuum is considered
as chromomagnetic superconducting medium~\cite{10}.
In this paper we consider the equation whose investigation
allowed the authors of~\cite{11} to come to the conclusion
on the existence of the solution having soft power
behaviour at the origin and study the general case of power
infrared behaviour with non-integer exponents.
Then we compare the obtained results with those
of~\cite{12,13}
dealing with the search for the field
equations solutions for the
propagator with non-integer power infrared behaviour.
Then we analyse the approximations made when obtaining
the closed equation for the propagator and discuss
the results.

The gluon propagator in
Baker - Ball - Zachariasen approach~\cite{5}
is defined by only one scalar function $Z(q^2)$,
\begin{eqnarray}
D_{\mu\nu}(q) &=& Z(q^2)D^{(0)}_{\mu\nu} (q), \;
D^{(0)}_{\mu\nu}(q) = -(q^2)^{-1}\Sigma_{\mu\nu}(q),\label{1}\\
\Sigma_{\mu\nu}(q)&=&g_{\mu\nu}-\frac{q_\mu\eta_\nu
+q_\nu\eta_\mu}{(q\eta)}
+\frac{q_\mu q_\nu \eta^2}{(q\eta)^2}.\label{2}
\end{eqnarray}
If we assume that the longitudinal part of triple gluon function
is dominant and neglect the transverse part then for $Z(q^2)$
the following equation can be obtained~\cite{5,7}:
\[
q^2\left(Z^{-1}(q^2)-1\right)(1-y) = 2^{-1}Z^{-1}(q^2)\int\frac{dk}{k^2k'^2}
\frac{(k\eta)-(k'\eta)}{\eta^2}\Sigma_{\lambda\sigma}(k)
\times
\]
\[
\times \Sigma_{\lambda\rho}(k')\left[Z(k'^2)(k\eta)
\rho(k,q)q_\sigma(q+k)_\rho\right.-
\]
\[
-Z(k^2)(k'\eta)\rho(k'q)q_\rho(q+k')_\sigma-Z(q^2)\left(Z(k'^2)(k\eta)-
Z(k^2)(k'\eta)\right)g_{\sigma\rho} -
\]
\be
-\left.Z(q^2)\rho(k,k')\left((k\eta)-(k'\eta)\right)\left((kk')g_{\sigma\rho}-
k_\rho k'_\sigma\right)\right]. \label{3}
\ee
In Eqs.~(\ref{1}) -- (\ref{3}) $\eta_\mu$ is gauge vector,
$k' = q-k$, $dk \equiv ig^2\mu^{4-n}Nd^nk/(2\pi)^n$, $n$ is
space-time dimension, $\rho (q_i,q_j)=(Z(q_i^2)-Z(q_j^2))
/(q_i^2-q_j^2)$.
Provided the condition $(q\eta)=0$ is imposed,
after a Wick rotation, approximate
angular integration~\cite{14} and the corresponding UV - subtractions,
the equation for renormalized
function $Z_R (q^2)=Z(-q^2)/Z(-\mu^2)$ has been obtained~\cite{11},
$\mu^2$ is renormalization point. This equation has the
following form:
\begin{eqnarray}
\frac{1}{Z_R(q^2)} &=& 1+\frac{3\alpha_s (\mu^2)}{\pi}
\left (T_1+\frac{T_2}{Z_R (q^2)}\right), \nonumber \\
T_1=\int\limits^{q^2}_{0}
F_1 (q^2,y)dy &-& \int\limits^{\mu^2}_{0}F_1 (\mu^2,y)dy
+ \int\limits^{\infty}_{q^2}F_2 (q^2,y)dy
-\int\limits^{\infty}_{\mu^2}F_2 (\mu^2,y)dy, \nonumber \\
T_2=\int\limits^{q^2}_{0}
F_4 (q^2,y)dy &-& \int\limits^{\mu^2}_{0}F_4 (\mu^2,y)dy
+ \int\limits^{\infty}_{q^2}F_5 (q^2,y)dy
-\int\limits^{\infty}_{\mu^2}F_5 (\mu^2,y)dy.  \label{4}
\end{eqnarray}
The functions $F_{1,2}(x,y)$ in Eq.~(\ref{4}) are linear in
$Z_R(x)$, $Z_R(y)$, $Z_R(x+y)$, and $F_{4,5}(x,y)$ are
quadratic.
As a result of numerical study~\cite{11}
the following parameterization was suggested to be used in
phenomenological applications
\begin{equation}
D(q^2)=\frac{Z_R(q^2)}{q^2}=\frac{m^{-2}}{c_1(q^2/m^2)^{\gamma_1}+
c_2(q^2/m^2)^{\gamma_2}+L(q^2/m^2)ln(\lambda_1 q^2/m^2+\lambda_2)}.
\label{5}
\end{equation}
It is essential that $\gamma_1 \simeq 0,22$ and
$\gamma_2 \simeq 0,86$ are non-integer parameters.
In Ref.~\cite{15} the possibility of the infrared behaviour
of the form
\begin{equation}
Z_R(x)=\left (x/\mu^2\right)^{1-c}
\left (\alpha_0 +\alpha_1x/\mu^2+\alpha_2
(x/\mu^2)^2+...\right) \label{6}
\end{equation}
was considered for the values
$0 < c < 1$ corresponding to the soft singular behaviour
of the propagator.
In Ref.~\cite{16} the  wider interval $-1 < c < 3$ was studied,
and in Ref.~\cite{17} the generalization was made
for the case of superposition of the terms with power infrared
behaviour,
\begin{equation}
Z^{-1}_{R}(x)=\sum^{N}_{n=1}
\left (x/\mu^2\right)^{c_n}f_n\left (x/\mu^2\right)
+f_0\left (x/\mu^2\right),\label{7}
\end{equation}
where functions $f_k(x/\mu^2), k=0,1,..., N$ are supposed to be
analytic at the origin and  non-integer parameters $c_n$ are
numbered  as
$c_1<c_2<...<c_N$ ($c_1=c-1$, $f_1(0)=$ $\alpha_0^{-1}$).
Equation~(\ref{4}) was shown  to have no
solutions
with infrared behaviour of  form~(\ref{6}), (\ref{7}),
if $-1<c<3$ and $c$ is non-integer (non-half-integer).
When  doing this, the technique  of extracting power terms
($\sim (q^2)^{-c_1}$,  $q^2\rightarrow 0$)
of the integral equation under consideration
defined only by the infrared behaviour of the propagator
 was developed.
In the infrared region  Eq.~(\ref{4}) takes the form:
\begin{eqnarray}
\frac{1}{Z_R(q^2)}&=&1+\frac{3\alpha_s(\mu^2)}{\pi}
\{P(q^2)+\frac{1}{Z_R(q^2)} Q(q^2)-\nonumber \\
&-&\alpha_0(q^2/\mu^2)^{-c_1}
\Delta (c)+o((q^2/\mu^2)^{-c_1})\},\label{8}
\end{eqnarray}
where $\Delta (c)$ is characteristic function,
 $P(q^2)$, $Q(q^2)$ are some regular at
$q^2=0$ functions which depend on the behaviour of $Z_R(q^2)$ at
all values of the argument.
At $c_1 > 0$ the term with the behaviour $(q^2)^{-c_1}$ is leading
at $q^2 \rightarrow 0$.
So, for the exponent we have the characteristic equation
 $\Delta (c)=0$.
At $c_1 <0$  the term  $(q^2)^{-c_1}$ in eq.~(\ref{8})
defined by the infrared region is not leading.
Nevertheless, the consideration of this case~\cite{15,17} shows
that for the infrared asymptotics~(\ref{6}) or~(\ref{7})
to be consistent with  Eq.~(\ref{4}) it is necessary
that $\Delta(c) = 0$.
It turns out that the function $\Delta(c)$ at $c=-1, -1/2, 0, 1/2,1,3$
has simple poles and in the intermediate points of the interval
$[-1,3]$ there are no roots of the equation $\Delta (c)=0$.
The explicit form of the characteristic function is
\begin{eqnarray}
\Delta(c)
&=&\frac{23}{24}\frac{1}{c+1}-\frac{187}{96}\frac{1}{2c+1}
+\frac{7}{24}\frac{1}{c}-\frac{5}{12}
\frac{1}{2c-1}+ \nonumber\\
&+&\frac{2}{3}\frac{1}{c-1}
-\frac{3}{4}\frac{1}{c-3}
+\tilde\Delta(c),\nonumber
\end{eqnarray}
\begin{eqnarray}
\tilde\Delta(c)&=&\frac{1}{24}\{
\int\limits^{1}_{0} dt [(1+t)^{1-c}
(28t^{-3+2c}+28t^{2-c}-9t^{-2+c}-
16(1+t^{-3+2c})\times \nonumber \\
&\times&(1-t^{1-c})/(1-t))-
 12t^{-3+2c} +4(1+3c)t^{-2+2c} + \nonumber \\
&+&2(16-5c-3c^2)t^{-1+2c} +
2(c^3+4c^2-13c+16)t^{2c}+ \nonumber \\
&+&1/6(192-122c+51c^2-22c^3-3c^4)t^{1+2c}-
16t^{1-c}+ \nonumber \\
&+&4(4c-15)t^{2-c}-
7t^{-2+c}+(7c-23)t^{-1+c}-(7c^2- \nonumber \\
&-&39c+64)t^c/2]-
23\gamma(2-c)+41\gamma(3-c)-11\gamma(4-c)+ \nonumber \\
&+&\gamma(5-c)+
\frac{6}{4-c}-\frac{1}{5-c}
+63/4-17c/6+7c^2/12+c^3/4\},\label{9}
\end{eqnarray}
where $\gamma(x)=(2^x-1)/x$. In Ref.~\cite{12} the approximate angular
integration in eq.~(\ref{3})
was not carried out and dimensional regularization was used for
which the ultraviolet subtractions are made automatically.
Denote by $\Delta_{dim}(c)$ the characteristic function of
Ref.~\cite{12} divided by $-2\Gamma(c)$ to pass to the
normalization of the present paper. It looks like:
\[
\Delta_{dim}(c)=\frac{\Gamma(2c-2)\Gamma^2 (3-c)}{2\Gamma^2 (c)
\Gamma(6-2c)}\{ \frac{2(5-2c)^2}{c-2}+\frac{1}{c-2}{}_3F_2(1,4-c,
2-c;c,7-2c;1)+
\]
\[
+\frac{2(c-1)}{c}{}_3F_2(1,3-c,2-c;c+1,7-2c;1)\}+
\frac{(12-5c)\Gamma(c-4)}{2\Gamma(c+1)}-
\]
\begin{equation}
-\frac{(c^2-6c+15)\Gamma(c-2)\Gamma^2(3-c)}{2\Gamma(c+1)
\Gamma(6-2c)}.
\label{10}
\end{equation}
At
$c<2$ the behaviour of the function $\Delta_{dim}(c)$
is similar to $\Delta(c)$
but
at $c>2$
the  behaviour of $\Delta_{dim}(c)$ qualitatively differs~\cite{17}
from that of $\Delta(c)$ and  at $c \simeq 2,537$ it
vanishes. This value corresponds to nearly oscillatory
potential $V(r) \sim r^{2,07}$.
However, we are not confident that this solution is consistent
for nonzero values of gauge parameter
$y = \gamma^{-1}=(p\eta)^2/p^2\eta^2$.

In Ref.~\cite{13} the equation for the gluon propagator in
simple loops approximation was investigated and self-consistency
equation for non-integer exponents was obtained.
Solutions of this equation were found but it remains unclear
whether its solutions correspond to the infrared asymptotics or
to the ultraviolet one.

Let us consider the assumptions made when obtaining the basic
equation~(\ref{3}) in some  detail.
The Schwinger -- Dyson equation for the gluon propagator in the
axial gauge has the following structure:
\begin{equation}
P_{\mu\nu}(p) - P^{(0)}_{\mu\nu}(p) = \Pi_{\mu\nu}(p)=\Pi^3_{\mu\nu}(p)
+\Pi^4_{\mu\nu}(p), \label{11}
\end{equation}
where $ \Pi^3_{\mu\nu}$ and $\Pi^4_{\mu\nu}$ are one-loop and
two-loop terms of polarization operator.
The inverse propagator $P_{\mu\nu}(p)$
depending on two scalar functions and free inverse propagator
$P^{(0)}_{\mu\nu}$ can be represented by
\[
P_{\mu\nu}(p)=P_aT^P_{\mu\nu}+P_bL_\mu L_\nu/p^2,\; P^{(0)}_{\mu\nu}=
-p^2T^P_{\mu\nu},
\]
where
\[
T^P_{\mu\nu}=g_{\mu\nu}-p_\mu p_\nu/p^2,\; L_\mu=p_\mu-\eta_\mu p^2/(p\eta).
\]
The tensor structure of one and two - loop terms is~\cite{9}:
\begin{eqnarray}
\Pi^3_{\mu\nu}(p)&=&\Pi^3_aT^P_{\mu\nu}+\Pi^3_bL_\mu L_\nu/p^2+\Pi^3_c
N_\mu L_\nu/(p\eta), \label{12} \\
\Pi^4_{\mu\nu}(p)&=&\Pi^4_bK_{\mu\nu}(p)+\Pi^4_cN_\mu L_\nu/(p\eta), \label{13}
\end{eqnarray}
where
\begin{eqnarray}
K_{\mu\nu}&=&\Sigma_{\mu\nu}-\gamma T^\eta_{\mu\nu}=(1-\gamma)T^P_{\mu\nu}+
L_\mu L_\nu/p^2, \nonumber\\
N_\mu&=&\eta_\mu-p_\mu\eta^2/(p\eta), \quad T^{\eta}_{\mu \nu}=g_{\mu \nu}-
\eta_{\mu} \eta_{\nu}/\eta^2.\label{14}
\end{eqnarray}
Note that the presence of nonsymmetric term in Eq.~(\ref{12})
makes it inconsistent to consider the tensor equation without
two-loop term. Let us see further what are the consequences of
the assumption on dominance of one tensor structure in the propagator,
$D_{\mu\nu}(p)=
Z(p)D^{(0)}_{\mu\nu}(p)$. In this case $P_a=-Z^{-1}(p)p^2$,
$P_b=0$ and tensor  Eq.~(\ref{11}) gives $\Pi^3_b=-\Pi^4_b$,
so we obtain the following tensor equation:
\begin{eqnarray}
P_{\mu\nu}(p)-P^{(0)}_{\mu\nu}(p)&=&\left(\Pi^3_a-(1-\gamma)
\Pi^3_b\right)T^P_{\mu\nu}=
\nonumber \\
&=&\frac{1}{1-y}\frac{\eta_\rho\eta_\lambda}{\eta^2}
\Pi^3_{\rho\lambda}T^P_{\mu\nu}. \label{15}
\end{eqnarray}
Consider the contribution of the tadpole term. We have
\begin{equation}
\Phi_{\mu\nu}\frac{\eta_\mu\eta_\nu}{\eta^2}=\frac{i}{(2\pi)^n}g^2N\int
d^nk\frac{Z(k)}{k^2}(n-2+\gamma). \label{16}
\end{equation}
It does not depend on the external momentum and it is some constant,
may be infinite.
Consider the assumption on independence of $Z$ of gauge parameter,
$Z(p)=Z(p^2)$.
This assumption has been shown~\cite{18} to be inconsistent
in the framework of the $BBZ$ approach~\cite{5}
and this approach should be modified.
Consider what  the consequences of the assumption for
the tadpole term are. The angular integral equals
\begin{equation}
\int d\Omega_n\left[\frac{k^2\eta^2}{(k\eta)^2}\right]=
\Omega_{n-1}\int_{-1}^{1}
(1-x^2)^{(n-3)/2}\left[\frac{1}{x^2}\right]dx=-(n-2)\Omega_n, \label{17}
\end{equation}
where $\Omega_n=2\pi^{n/2}/\Gamma(n/2)$, and the singularity
at the origin is understood in any of the senses:
\[
\left[\frac{1}{x^2}\right]=PV\frac{1}{x^2},\;
\frac{1}{(x+i0)^2},\; \frac{1}{(x-i0)^2},
\; |x|^\lambda\mid_{\lambda=-2}=x^{-2}.
\]
So, for reasonable axial singularity prescriptions and
sensible regularizations (e.g., dimensional regularization,
symmetric integration) the tadpole term vanishes for $Z(p)=Z(p^2)$.
For $Z(k^2) \mid_{k^2 \rightarrow \infty}
\sim ln^\alpha k^2$
the divergent term in Eq.~(\ref{3}) of the form
$\sim \Lambda^2ln^\alpha\Lambda^2$
should be subtracted by the corresponding mass counterterm.
The characteristic equations have no solutions at
 $c=2$ $(D\sim 1/(q^2)^2)$.
In this case the infrared contribution is a constant and it could
be considered as a part of mass renormalization constant~\cite{18}.
Before doing this let us look more attentively at the condition
$(p\eta)=0$
imposed to make the problem tractable. As it was noted in~\cite{19},
the Poincar\`e-Bertrand singular terms should be taken into account
when using the decomposition formula for the axial denominators.
For generalized $c$-prescription~\cite{20}
\begin{equation}
\left[\frac{1}{(k\eta)}\right]=\frac{1}{(k\eta)+i0}+c\delta((k\eta)) \label{18}
\end{equation}
the decomposition formula has the form:
\begin{eqnarray}
\left[\frac{1}{(k\eta)}\right]\left[\frac{1}{(p-k,\eta)}\right]&=&
\left[\frac{1}{(p\eta)}\right]
\left(\left[\frac{1}{(k\eta)}\right]+\left[\frac{1}{(p-k,\eta)}
\right]\right)+ \nonumber \\
&+&c(2\pi i-c)\delta\left((k\eta)\right)\delta\left((p\eta)\right). \label{19}
\end{eqnarray}
At $c=0,2\pi i$ the Poincar\`e-Bertrand terms are absent and in the case
$c=i\pi$ corresponding to principal value prescription, one should
account for them. Consider whether these terms prevent from going to the limit
$(p\eta)=0$
in the integral equation~(\ref{3}).
It can be seen that in product of the axial singularities we can not
go to the limit
 $(p\eta)=0$,
\begin{equation}
\left.PV\frac{1}{(k\eta)}PV\frac{1}{(p-k,\eta)}\right|_{(p\eta)\rightarrow 0}
\rightarrow-PV\frac{1}{(k\eta)^2}-\pi^2\delta\left((k\eta)\right)\delta
\left((p\eta)\right).
\label{20}
\end{equation}
It can be shown that for $Z(q^2)=-M^2/q^2$
the additional contribution in Eq.~(\ref{3}) has the form:
\[
\Delta_{PB}(p)=g^2M^2N(p^2/\mu^2)^\epsilon\sqrt{p^2\eta^2}\delta((p\eta))
\times
\]
\be
\times\{y^{-1/2+\epsilon}[\phi_1(\epsilon)+y\phi_2(\epsilon)+O(y^2)]
+\chi_1(\epsilon)+y\chi_2(\epsilon)+O(y^2)\}, \label{21}
\ee
and $\phi_1(\epsilon)=0, \chi_1(\epsilon)\mid_{\epsilon\rightarrow0}
\sim\epsilon$, $\epsilon=(n-4)/2$.
At $\epsilon=0$
\[
\Delta_{PB}(p)=\delta((p\eta))\sum_{k=1}c_k(p\eta)^k=0,
\]
and the Poincar\`e-Bertrand terms do not prevent from going to the limit
$(p\eta)\rightarrow0$.
It is reasonable to assume that this is true
for the less singular behaviour of $Z(k^2)$ at zero.

Consider the results for the case
$(p\eta)\not = 0$. For $Z(q^2)=-M^2/q^2$
it was obtained in Refs.~\cite{8,9,21} that,
\begin{equation}
\Pi^3_{\mu\nu}\frac{\eta_\mu\eta_\nu}{\eta^2}=-\frac{g^2M^2N}{16\pi^2}
[3y-3-yln(4y)-F(y)/2], \label{22}
\end{equation}
where
\[
F(y)=y\int_{0}^{1}ln(ty)(1-t)^{-1/2}(1-ty)^{-1}dt.
\]
If one keeps the dependence on $\epsilon$, then  at $y\rightarrow0$
the result can be represented as a series of the form
\begin{equation}
\Pi^3_{\mu\nu}\frac{\eta_\mu\eta_\nu}{\eta^2}=-\frac{g^2M^2N}{16\pi^2}
[7-10y^\epsilon+a_1(\epsilon)y+b_1(\epsilon)y^{1+\epsilon}+\cdots].
\label{23}
\end{equation}
At $\epsilon<0$ it is impossible to go to the limit $y\rightarrow0$,
and  at $\epsilon>0$ one can do it with the result 7 in brackets of
Eq.~(\ref{23}).
If one puts $\epsilon=0$ at first, then  $-3$ in brackets of
Eq.~(\ref{23}) at
$y=0$ is obtained. Thus the limits $y\rightarrow0$ and $\epsilon\rightarrow0$
do not commute and the point $(p\eta)=0$
is singular for  Eq.~(\ref{3}) at least for $Z(q^2)\sim1/q^2$.
Having the dependence on the gauge parameter $y$ in~(\ref{22}),
it is impossible to solve  Eq.~(\ref{3}) asymptotically
by mass subtraction.

Let us verify the correctness of the assumption on dominance of
longitudinal part of the vertex function. The direct
calculation~\cite{8,9} shows that the transverse part
of triple gluon vertex  $\Gamma^T$
introduced in ref.~\cite{7} of the form
\[
i\Gamma^T_{3\lambda\mu\nu}(p,q,r)=\xi M^{-2}\{g_{\lambda\mu}[p_\nu(qr)-
q_\nu(pr)]+
\]
\be
+g_{\mu\nu}[q_\lambda(pr)-r_\lambda(pq)] +  g_{\nu\lambda}
[r_\mu(pq)-p_\mu(qr)] + r_\lambda p_\mu q_\nu - q_\lambda r_\mu p_\nu \}
\label{24}
\ee
gives nonzero contribution which completely cancels
$y$-dependent right hand side of Eq.~(\ref{22}) at $\xi=1$.
So, taking into account the transverse part
of triple gluon vertex (which is not fixed
by Slavnov - Taylor identities)
it turns out possible to satisfy asymptotically the Schwinger - Dyson
equation for the propagator with infrared behaviour
$D(q)\sim1/(q^2)^2,
q^2 \rightarrow0$.
Note that the transverse part of vertex function~(\ref{24})
naturally meets the scaling property~\cite{21}.
It seems unlikely that there are other ways to cancel
the leading at $q^2\rightarrow 0$ term of the Schwinger - Dyson
equation which has the complicated dependence on the gauge parameter.

We conclude that the power type infrared behaviour~(\ref{6}), (\ref{7})
of the gluon propagator is inconsistent,
whereas the behaviour
$D(q)\sim1/(q^2)^2, q^2 \rightarrow 0 $ is consistent.
Despite of the differences in the argumentation, the same
conclusion was made by the authors of the recent paper~\cite{22}.

I would like to thank B.~A.~Arbuzov, R.~Oehme,
V.~E.~Rochev, A.~A.~Slav\-nov for helpful discussions.
I am most grateful to L.L. Jenkovszky and all members of the
Organizing Committee for a kind invitation and hospitality.
This research was supported in part by the Russian Foundation of
Fundamental Investigations, Grant No.~95-02-03704-a.

\end{document}